# Detecting Dynamic States of Temporal Networks Using Connection Series Tensors

Shun Cao[1,2], Hiroki Sayama[1,2,3]

[1] Center for Collective Dynamics of Complex Systems, Binghamton University, State University of New York, Binghamton, NY 13902-6000, USA

[2] Department of Systems Science and Industrial Engineering, Binghamton University, State University of New York, Binghamton, NY 13902-6000, USA

[3] Waseda Innovation Lab, Waseda University, Shinjuku, Tokyo 169-8050, Japan

Correspondence should be addressed to Shun Cao: scao12@binghamton.edu

## Abstract

Many temporal networks exhibit multiple system states, such as weekday and weekend patterns in social contact networks. The detection of such distinct states in temporal network data has recently been explored as it helps reveal underlying dynamical processes. A commonly used method is network aggregation over a time window, which aggregates a subsequence of multiple network snapshots into one static network. This method, however, necessarily discards temporal dynamics within the time window. Here we develop a new method for detecting *dynamic* states in temporal networks using information regarding the timeline of contacts between each pair of nodes. We apply a similarity measure informed by the techniques of processing time series and community detection to sequentially discompose a given temporal network into multiple dynamic states (including repeated ones). Experiments with empirical temporal network data demonstrated that our method outperformed the conventional approach using simple network aggregation in revealing interpretable system states. In addition, our method allows users to analyze hierarchical temporal structures and to uncover dynamic state at different spatial/temporal resolutions.

## 1. Introduction

Temporal networks are a useful framework to represent and analyze time-dependent changes and underlying dynamics of complex systems [1,2,3]. Many phenomena, ranging from disease spread [4,5,6] and human communication [7,8,9] to financial transactions [10,11] and human brains [12,13], can generate large-scale temporal network data. In many cases, the temporal network data can often be broken down into a sequence of discrete system states, some of which may reoccur many times. For example, air traffic networks can show seasonal variations [14,15] and peak/off-peak weekly patterns [15], which can be modeled and studied as a temporal sequence of distinct system states. System state detection in temporal networks is useful for investigating the dynamics of time-varying complex systems and making better interpretation of large-scale temporal network data sets.

To detect the system states in temporal networks, Masuda and Holme recently proposed an approach using network aggregation and graph similarity [16]. Their method first partitioned a given temporal network into a sequence of static networks through network aggregation. Then they measured the graph similarity among the aggregated static networks to generate a distance matrix. Further, they applied hierarchical clustering and Dunn's index [17] to obtain the final system states. In their method, the timelines of contacts between pairs of nodes within a time window were aggregated as the static edge weights in the corresponding static networks. Yet, these timelines of contacts may offer critical information in exploring the system dynamics of temporal networks. For example, the patterns regarding the fluctuation of dynamic contacts between distinct regions in brain networks can indicate various brain activities or states [18,19,20]. Here, we use *connection series* to define the timeline of contacts between a pair of nodes. It refers to the timeline of *connection status* (e.g., connected and disconnected) between two nodes, which has a structure of a bit vector (i.e., bit map, bit set, or bit string) consisting only elements of "1" (the status of connected) and "0" (the status of disconnected). Figure 1 gives a simple example of the connection series between two nodes. Figure 2 plots the two real-world connection series regarding face-to-face contacts between two students in a primary school [21] and that between two attendees at an academic conference [22], respectively, in which we can observe distinct fluctuation patterns between each pair of individuals over time. We downloaded both empirical data sets from SocioPattern.org.

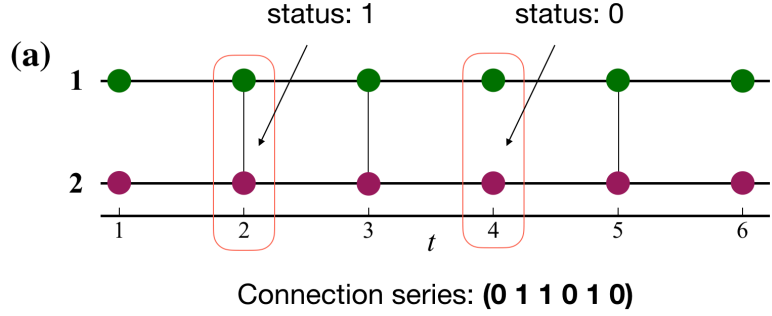

Figure 1: A simple example of connection series between two nodes.

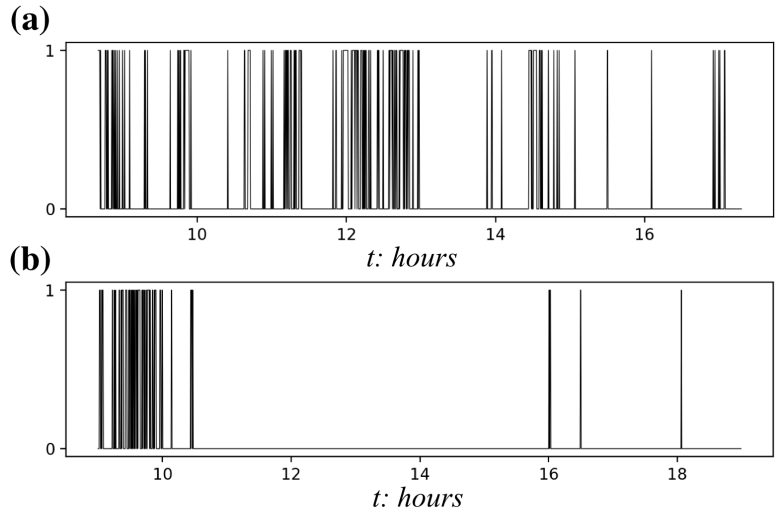

Figure 2: Real-world connection series between two nodes. (**a**) a face-to-face connection series involving "node 1558" and "node 1567" in the first day's primary school data [21]. (**b**) a connection series involving "node 1080" and "node 1125" in the first day's conference data [22]. 1 and 0 in *y*-axis represent two individuals interact (are connected) and not interact (are disconnected) with each other, respectively, at corresponding timestamps. Note that the sampling interval (i.e., time resolution) is 20 seconds in both (a) and (b).

In this study, we propose a method to detect dynamic states in temporal networks using connection series between nodes. Like in [16], we divide a given temporal network data into subsequences consisting of network snapshots using non-overlapping time windows. Our method then transforms each subsequence into a tensor, in which every element is the connection series between the corresponding pair of nodes. These tensors generated from multiple subsequences are then linked with one another and form a meta-level network whose edge weights are similarities between them. Non-overlapping communities are detected on this meta-level network to classify each subsequence into one of the distinct dynamic states (represented as communities in the meta-level network). Experiments using two empirical temporal network data sets demonstrated that our method was capable of detecting interpretable and practical dynamic states in temporal networks. Additionally, by comparing the results with the already known sequence of events that took place in each data set, our method also outperformed the previous approach. The temporal networks can be undirected or directed. Not that we only consider the unweighted temporal networks here.

We organized the rest of this paper as follows. Section 2 describes our proposed method. Section 3 describes empirical temporal network data sets used in experiments. Section 4 presents the results. In Section 5, we conclude and discuss limitations.

## 2. Method

We transform or represent a temporal network data set as a temporal network with $n$ network snapshots, $S_O = [G_{t_1}(V_{t_1}, E_{t_1}), ..., G_{t_n}(V_{t_n}, E_{t_n})]$, where $G_{t_i}$ is the network snapshot at timestamp $t_i$, in which $V_{t_i}$ and $E_{t_i}$ denote the node set and edge set, respectively. In this representation, the timeline is discrete and $t_{i+1} = t_i + \Delta t$, where $\Delta t$ is the sampling interval of the original temporal network data set which is also known as time resolution. Note that, almost all empirical temporal network data sets are of the form of contact lists, i.e. the two nodes involved and the time of the interaction, in which those timestamps without any events are usually not recorded. We complete each unrecorded timestamp with a synthesized network snapshot, whose node set is defined the same as its nearest recorded network snapshot prior to it, while the set of edges is made up of all zeros. This simple data processing approach enables our method to be applicable to most of the temporal network data sets.

Figure 3 presents a schematic overview of our method. First, we split the whole temporal network data into $T$ subsequences using nonoverlapping time windows of length $w \ll n$. The last subsequence may be shorter than $w$ since $n$ can be not divisible by $w$. We denote these subsequences as $S_S = [s^1, ..., s^T]$. Second, we transform each subsequence into a connection series tensor $A^{(i)}$, whose structure is similar to the adjacency matrix of a network, in which each element is not a value but a connection series between each pair of nodes that appears in this subsequence. We denote these obtained connection series tensors as $A = [A^{(1)}, ..., A^{(T)}]$, where $A^{(i)}$ is the $i_{th}$ tensor. The format of a connection series tensor $A^{(i)}$ is shown as below.

$$A^{(i)} = \begin{bmatrix} c_{1,1}^{(i)} & c_{1,2}^{(i)} & \cdots & c_{1,V^{(i)}}^{(i)} \\ c_{2,1}^{(i)} & c_{2,2}^{(i)} & \cdots & c_{2,V^{(i)}}^{(i)} \\ \vdots & \vdots & \vdots & \vdots \\ c_{V^{(i)},1}^{(i)} & c_{V^{(i)},2}^{(i)} & \cdots & c_{V^{(i)},V^{(i)}}^{(i)} \end{bmatrix}, \quad (1)$$

where $c_{j,k}^{(i)}$ denotes the connection series between nodes $j$ and node $k$, while $V^{(i)}$ is the set of nodes that appears in the subsequence $s^i$. Note that we do not apply the node set of the whole temporal network $V_{all} = V^{(1)} \cup ... \cup V^{(T)}$ to each connection series tensor since $V^{(i)}$ is generally much smaller than $V_{all}$, which can help avoid low-efficiency computation in the rest steps. Third, we quantify the similarity between every pair of these connection series tensors by a measure that will be described later. Forth, we construct a fully connected, weighted meta-level network whose nodes and edges represent these tensors and the similarities between them, respectively. Finally, we run the Louvain method [23] on the meta-level network to classify these connection series tensors (nodes in this meta-level network) into multiple communities that are interpreted as dynamic states. We can also adjust the community resolution in the Louvain method, a tunable parameter, to study dynamic states at different spatial/temporal resolutions in a given temporal network.

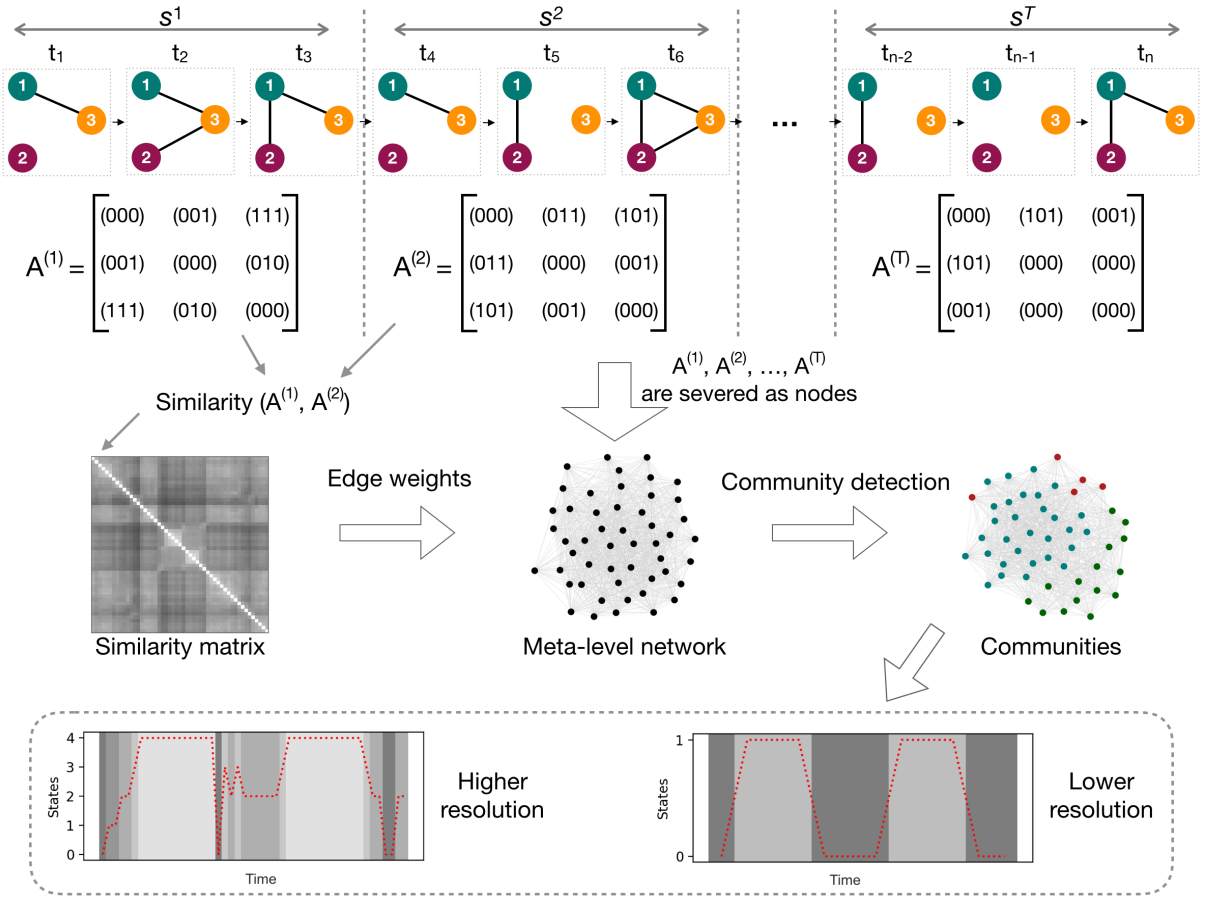

Figure 3: A schematic overview of our proposed method. Here we present a temporal network made of three nodes and set $w = 3$.

Similarity between Connection Series Tensors

Let the two connection series tensors we compare be $A^{(a)}$ and $A^{(b)}$ ($a \neq b$), whose node sets are $V^{(a)}$ and $V^{(b)}$ respectively. The format of $A^{(a)}$ and $A^{(b)}$ may be different due to probably distinction of $V^{(a)}$ and $V^{(b)}$. To make the format of $A^{(a)}$ and $A^{(b)}$ consistent, we transform $A^{(a)}$ and $A^{(b)}$ into $A'^{(a)}$ and $A'^{(b)}$, respectively, both of whose node sets are redefined as $V'^{(a)} = V'^{(b)} = V^{(a)} \cup V^{(b)}$. In this case, a few nodes may be stay at the disconnected status during the subsequence of $s^{(a)}$ or $s^{(b)}$, which leads the corresponding connection series in $A'^{(a)}$ and $A'^{(b)}$ to be made up of all zeros. The steps of our proposed similarity measure are:

*Step 1*. We compute the similarity between every pairwise connection series in $A'^{(a)}$ and $A'^{(b)}$, say $sim(c_{j,k}^{(a)}, c_{j,k}^{(b)})$.

*Step 2*. We average all the similarities obtained in step 1 as the similarity between $A^{(a)}$ and $A^{(b)}$. The formula is shown as below.

$$Sim(A^{(a)}, A^{(b)}) = Sim(A'^{(a)}, A'^{(b)}) = \frac{1}{m(m-1)} \times \sum_{j}^{m} \sum_{k!=j}^{m} sim(c_{j,k}^{(a)}, c_{j,k}^{(b)}), \qquad (2)$$

where $m$ denotes the number of nodes in $V'^{(a)}$ and $V'^{(b)}$. Note that we exclude all self-connection series $c_{j,j}^{(i)}$ to save a little computational resource, because $c_{j,j}^{(i)}$ only consists zeros.

To compute the similarity between two connection series, $sim(c_{j,k}^{(a)}, c_{j,k}^{(b)})$, we develop an simple method that was informed by the well-developed similarity measures of time series [24,25,26]. Our similarity measure is computed based on the best match between tow series. A schematic illustration of our proposed method is shown in Figure 4. We first generate multiple matching cases between $c_{j,k}^{(a)}$ and $c_{j,k}^{(b)}$ by rolling and rotating. Then we pinpoint the case with the maximum number of corresponding elements that equals one another (we will call it "matched elements" in the rest of the paper) in $c_{j,k}^{(a)}$ and $c_{j,k}^{(b)}$. Note that, the elements refer to those connection status (e.g., "0" and "1") in connection series. Finally, we compute $sim(c_{j,k}^{(a)}, c_{j,k}^{(b)})$ using the formula below.

$$sim\left(c_{j,k}^{(a)}, c_{j,k}^{(b)}\right) = \frac{\max_{1 \leq p \leq w}(M_1, M_2, \ldots M_w)}{w}, \qquad (3)$$

where $M_i$ represents the number of matched elements in matching case $i$, while $w$ is length of time window. Note that the maximum length of the connection series of $c_{j,k}^{(a)}$ and $c_{j,k}^{(b)}$ equals the length of time window $w$. This similarity can be seen as the maximum matching ratio of two given connection series.

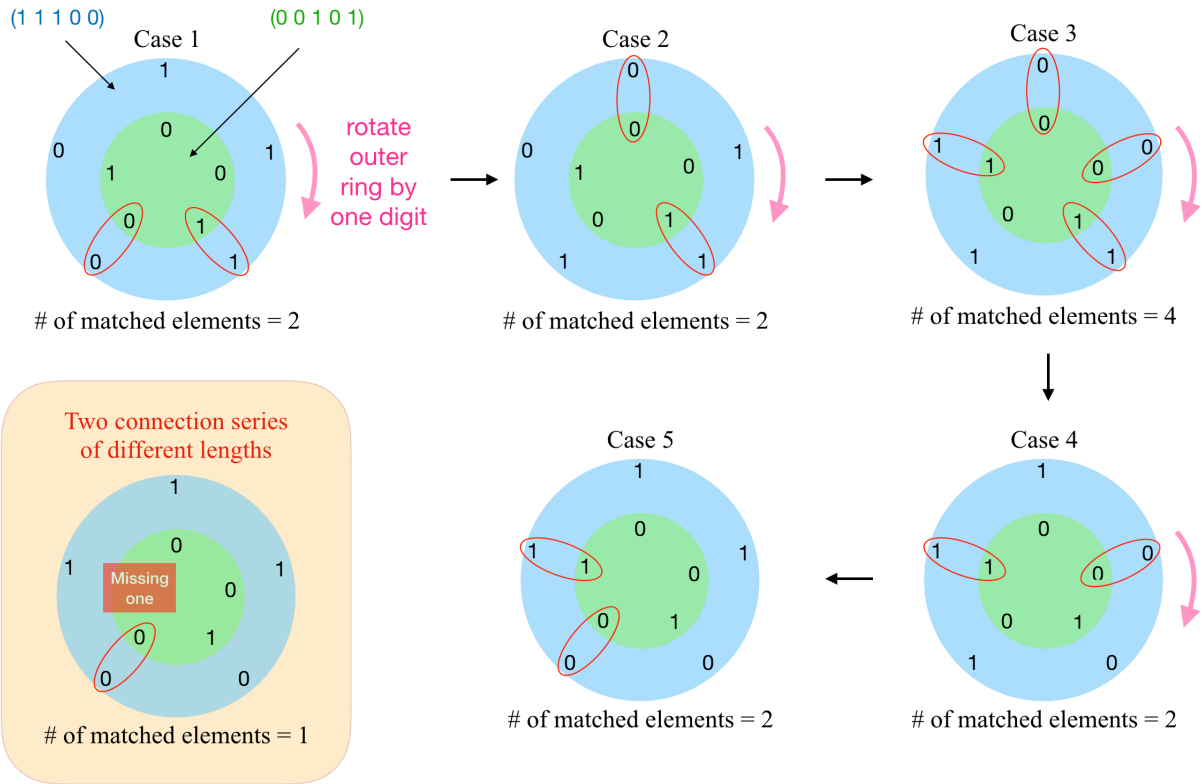

Figure 4: A schematic diagram of our proposed similarity measure for connection series. Given two connection series, say (1 1 1 0 0) and (0 0 1 0 1), we keep the sequential order of each of the series unchanged and find the best matching between them. To do this, we roll both of the connection series into rings, fix one of them, say (0,0,1,0,1), as the inner ring, and the other one, say (1,1,1,0,0), as the outer ring, respectively. Then we rotate the outer ring clockwise element by element until we make a full rotation and obtain multiple matching cases regarding one-to-one correspondence between elements in the respective rings (i.e., series). In this example, we have five matching cases. Note that the number of matching cases is equal to the length of connection series (i.e., length of time window). Then we count the number of matched elements in each case and choose the one that has the most matched elements as the best matching between these two given connection series. Here, case 3 provides the best matching between (1,1,1,0,0) and (0,0,1,0,1), in which the number of matched elements is 4. Finally, we divide this number by the length connection series $w$ (length of the window), where $w = 5$ in the example shown here. According to formula (3), the final similarity between (1,1,1,0,0) and (0,0,1,0,1) is 0.8. The bottom left panel in a light-yellow frame gives an example when two connection series are different in length, say (1,1,1,0,0) and (0,0,1,0). In this case, we fix the shorter one as the inner ring and the longer one as the outer ring, respectively. Then we still keep a clockwise-moving way that is based on the one-to-one correspondence of elements in the two respective rings and leave the places with no elements empty in the shorter ring (the red square in the bottom left panel). The rest process of finding the best matching case and computing similarity is the same as previously.

Community Detection on Meta-level Network

We apply community detection to the generated meta-level network to assign each node (= a connection series tensor, or a subsequence of the original temporal network) with a distinct dynamic state label. Many community detection algorithms have been developed and employed with varying levels of success [23,27,28,29]. Here we use the Louvain method [23], one of the most popular modularity maximization algorithms. The Louvain method is beneficial to our work for two main reasons. First, it can take edge weights into account. Two nodes connected by an edge with greater edge weight (i.e., the higher similarity between connection series tensors) in the meta-level network are more likely to be assigned to the same dynamic state. Second, it also provides a tunable parameter of community resolution that allows for the exploration of dynamic states at different spatial/temporal resolutions of interest, which is especially helpful for unknown temporal networks.

## 3. Data

We used the primary school and conference data sets that were downloaded from SocioPattern.org to run experiments. Rather than arbitrarily selecting empirical data sets, we chose them because there were known "ground truth" states to evaluate the performance of our method. Both the data sets represent the physical proximity between people. Table 1 lists the basic properties of the two data sets.

Table 1: The properties of the temporal network data sets used in this paper.

| Data | Primary school data [21] | Conference data [22] |
|---|---|---|
| Number of nodes | 242 | 110 |
| Selected Period | First day, 8:40 ~ 17:20 | First day, 9:00 ~ 19:00 |
| Sampling interval | 20 seconds | 20 seconds |

Primary School Data

The primary school data was collected in a primary school in Lyon, France. In this school, each of all the five grades was divided into two classes [21]. The schedule of a school day was shown in Table 2. Different classes took turns to take breaks in a playground and to have lunch in a canteen because the playground or the canteen could not accommodate all the students at the same time [21]. The face-to-face contacts between 232 children and 10 teachers in the school were measured and recorded by body-mounted RFID devices. Two individuals were joined when they faced each other in a close range (about 1 m to 1.5 m). The data were collected from 8:45 to 17:20 on Thursday, October 1st, 2009, and from 8:30 to 17:05 on Friday, October 2nd, 2009. We used only the first day's data in this paper.

Table 2: Schedule of a school day in the primary school in Lyon, France [21].

| Time | Event |
|---|---|
| 8:30 ~ 10:30 | Class time |
| 10:30 ~ 10:55 (approximate time) | Break time |
| 10:55 ~ 12:00 | Class time |
| 12:00 ~ 14:00 | Lunchtime |
| 14:00 ~ 15:30 | Class time |
| 15:30 ~ 15:55 (approximate time) | Break time |
| 15:55 ~ 16:30 | Class time |

Conference Data

This data set is named "Hypertext 2009 dynamic contact network" on the website of SocioPattern.org. Here we call it "conference data" for short. The data set represents the temporal network of face-to-face contacts of about 110 attendees at an academic conference. It was collected during the ACM Hypertext 2009 conference (http://www.ht2009.org/) hosted by the Institute for Scientific Interchange Foundation in Turin, Italy, from June 29th to July 1st, 2009 [22]. The data collection method was the same as that used for the primary school data. We used only the first day's (Monday, June 29, 2009) data in this paper, whose program was given in Table 3.

Table 3: The first day's program of ACM hypertext 2009 conference (http://www.ht2009.org/).

| Time | Event |
|---|---|
| 9:00 ~ 10:30 | Set-up time for posters and demos |
| 10:30 ~ 11:45 | Workshops 1 |
| 11:45 ~ 12:00 | Coffee break 1 |
| 12:00 ~ 13:30 | Workshops 2 |
| 13:30 ~ 15:00 | Lunch break |
| 15:00 ~ 16:30 | Workshops 3 |
| 16:30 ~ 16:45 | Coffee break 2 |
| 16:45 ~ 18:05 | Workshops 4 |
| 18:05 ~ 18:10 | Short break |
| 18:10 ~ 19:00 | Wine & cheese welcome reception |

## 4. Experiments

Here, we applied our proposed method to the two real-world temporal networks to demonstrate how this approach can be used to detect meaningful insights regarding complex interactions among elements in time-varying complex systems. We used the event information shown in Tables 2 and 3 as the ground truth for our results. In the experiments, we varied the community resolution parameter in the Louvain method from 1.0 to smaller values (decreasing 0.01 in each variation) to scan the hierarchical temporal structure and uncover dynamic states at different resolutions. Note that smaller community resolution parameter in the Louvain method indicates a higher resolution of system states and vice versa. For comparison, we also implemented the approach using network aggregation and graph similarity that proposed in [16]. Here we chose DeltaCon [30] form multiple graph similarity measures used in [16].

Results for Primary School Data

We partitioned the primary school data into subsequences through time windows of length 20 minutes, which was the same as that used for the same data set in [16]. Figure 5 presents the results for the primary school data obtained by our method. Figure 5(b) exhibits the detected two dynamic states using the community resolution parameter of 1.0, state 0 (the approximated periods are 8:40 ~ 11:50 and 14:10 ~ 17:20) and state 1 (the approximated period is 11:50 ~ 14:10). By comparing the results with the primary school's schedule, it is reasonable to correspond states 1 and 0 to the lunchtime and class time, respectively. However, the morning and afternoon breaks were not revealed at such a low resolution of system states.

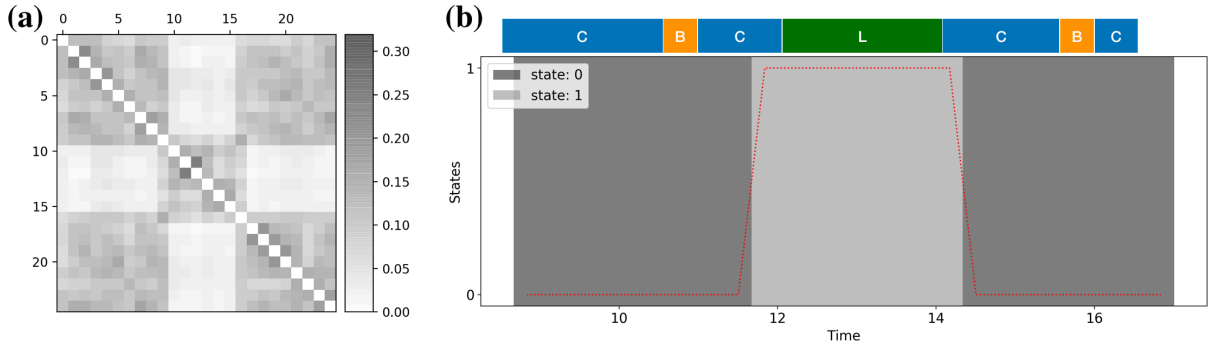

Figure 5: Results for the primary school data using our proposed method. (**a**) Similarity matrix. (**b**) Detected states using the community resolution parameter of 1.0. We use the red dotted line and spans with distinct greyscales to represent the dynamic states. The horizontal bar on the top of figure (b) is the schedule of a school day. The capital letter 'C', 'B', and 'L' in the bar correspond to the class time, the break time, and the lunchtime in the schedule, respectively. Note that the period (8:45 ~ 17:20) of the collected data does not match with that of a school day's schedule (8:30 ~ 16:30).

Figure 6 provides more representative results using distinct community resolutions parameters. The results obtained by setting the community resolution parameter as 0.92 are shown in Figure 6(b). There are three detected dynamic states, state 0, state 1, and state 2 which are consistent with class time, break time (morning and afternoon), and lunchtime, respectively. The detected two breaks are longer than the break time in the schedule of a school day, which may be led by those students in different classes took turns to take breaks due to the limitation of the playground [21]. Figure 6(c) gives the results that are obtained by using an even smaller community resolution parameter of 0.91. It shows that the morning break can be broken into two dynamic states, states 1 and 2, which may suggest a distinction in collective behaviors of different classes at the morning break time. If we continue to decrease the community resolution parameter to 0.89 (Figure 6(d)), we can even find the subtle differences between morning class time (state 0 in Figure 6(d)) and afternoon class time (state 4 in Figure 6(d)). Additionally, the results in Figure 6 indicate that smaller community resolution parameter (high resolution of system states) is beneficial to discover more subtle dynamic states.

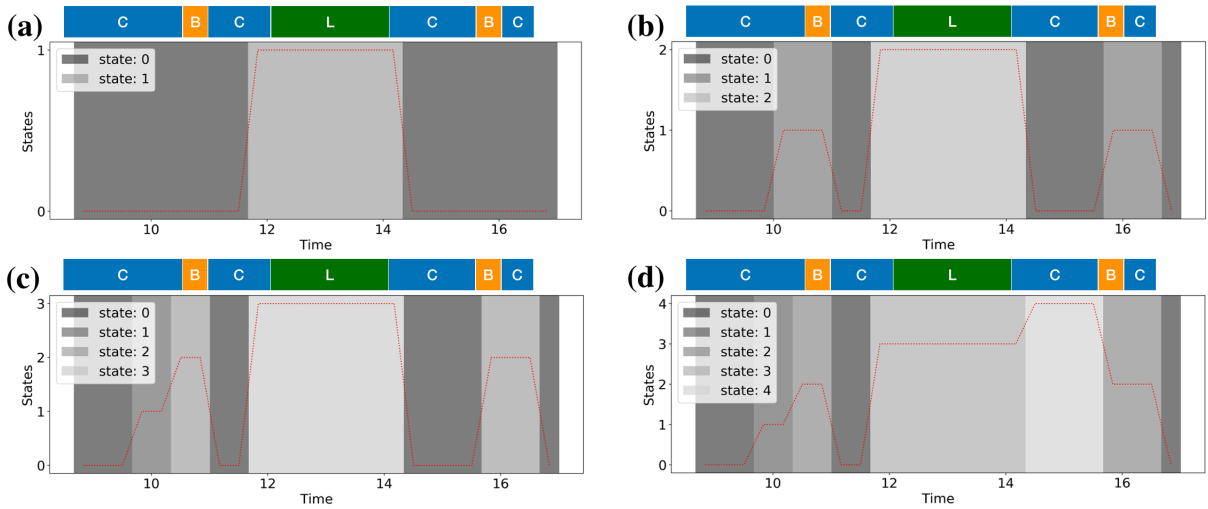

Figure 6: More representative results for the primary school data using our proposed method. (**a**) Detected states obtained by setting the community resolution parameter as 1.0, which is the same as the results in Figure 4(b). (**b**) Detected states using the community resolution parameter of 0.92. (**c**) Detected states obtained by setting the community resolution parameter as 0.91. (**d**) Detected states using the community resolution parameter of 0.89. Note that the period (8:45 ~ 17:20) of the collected data does not match with the schedule of a school day (8:30 ~ 16:30).

As a comparison, Figure 7 gives the results for the primary school data obtained by the method using network aggregation and graph similarity in [16]. This method discovered two optimal system states shown in Figure 7(b) through hierarchical clustering and Dunn's index [17]. Though it detected the lunchtime and class time, it failed to recognize the two breaks between classes.

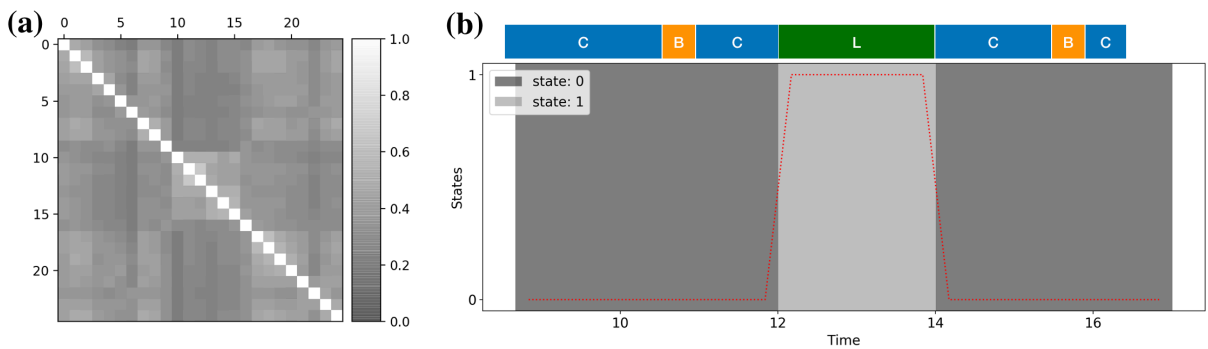

Figure 7: The results for the primary school data obtained by the method using network aggregation and graph similarity in [16]. (**a**) Similarity matrix. (**b**) Detected system states. Note that the period (8:45 ~ 17:20) of the collected data does not match with the schedule of a school day (8:30 ~ 16:30).

Results for Conference Data

The conference data was divided into subsequences by nonoverlapping time windows of length 5 minutes, which are chosen according to the shortest event (short break (18:05 ~ 18:10) in Table 3). Figure 8 shows the results obtained by our proposed method, which are mainly aligned with the first day's program of ACM hypertext 2009 conference. For example, the results in Figure 8(b) (community resolution parameter equals 1.0) suggest that coffee break 1, coffee break 2, wine & cheese welcome reception, and part of lunch break (approximately from 13:30 ~ 15:15) are all recognized as the same dynamic state 2. Workshops 1, 2, 3, and 4 correspond to state 0, state 3, state 1, and state 4, respectively. Figure 8(c) provides more subtle insights using a smaller community resolution parameter (0.99), in which state 5 emerges at the end of the whole day's program. This new state may be interpreted as a "ending time" of the first conference day.

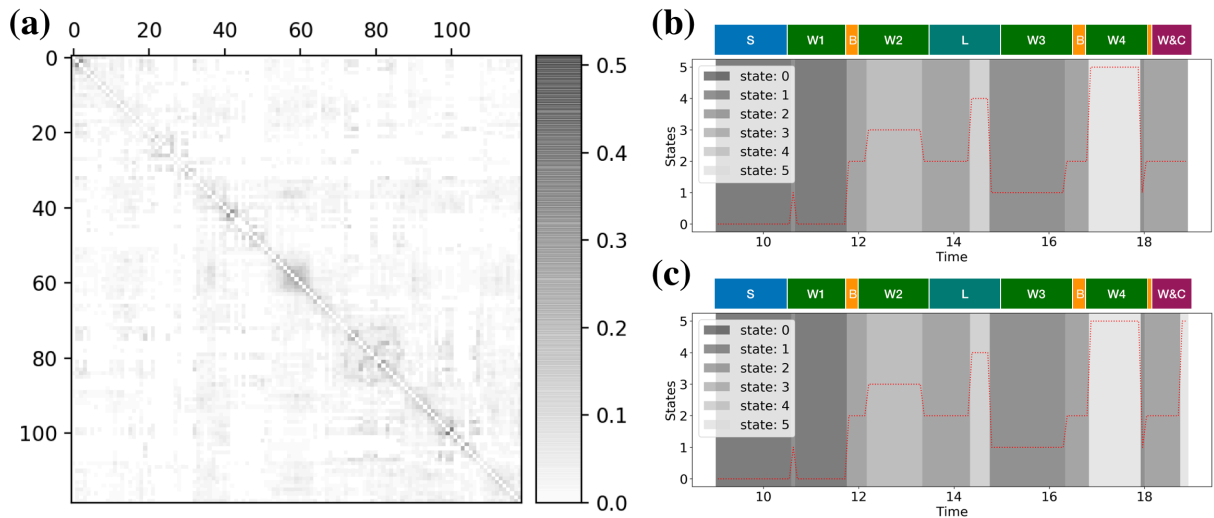

Figure 8: Results of the conference data using our proposed method. (**a**) Similarity matrix. (**b**) Detected states using the community resolution parameter of 1.0. (**c**) Detected states using the community resolution parameter of 0.99. The horizontal bar on the top of figure (b) and (c) is the first day's program of ACM hypertext 2009 conference. The capital letter 'S', 'W', 'B', 'L', and 'W&C' in the bar correspond to the set-up time, the workshop, coffee break, lunchtime, and wine & cheese welcome reception in the first day's program, respectively.

On the contrary, the results obtained by using the method in [16] performed poorly in the conference data. As shown in Figure 9(b), the only two detected system states, states 0 and state 1 failed to match the first day's conference program.

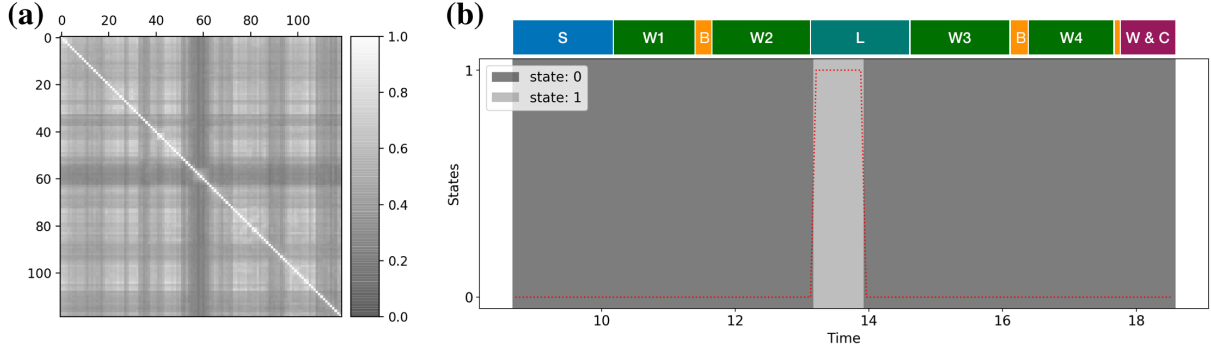

Figure 9: Results for the conference data using the method in [16]. (**a**) Similarity matrix. (**b**) Detected system states.

## 5. Conclusions

In this paper, we developed a new method of detecting dynamic states in temporal networks. We transformed a given temporal network into a sequence of tensors that consisted of connection series between each pair of nodes. These connection series can help capture the collective dynamics regarding temporal and spatial interactions between elements in time-varying complex systems. We also proposed a simple method to evaluate the similarity between two connection series tensors, which can be potentially extended to a similarity measure of two temporal networks. The results for empirical temporal networks demonstrated the effectiveness of our method in detecting dynamic states. Our method also outperformed the approach proposed by Masuda and Holme [16] in revealing actual events in real-world temporal networks, which suggests that incorporating timelines of contacts between pairs of nodes within time windows helps system state detection. Users can use our method to break down a large temporal network into a sequence of discrete dynamic states and scan state dynamics at various spatial/temporal resolutions. Moreover, they can investigate the temporal or topological properties in certain dynamic states of interest.

Meanwhile, our work has a few limitations. First, we arbitrarily choose the size of the nonoverlapping time window, which may need more discussions in the context of state detection. Second, the proposed similarity measure for the connection series tensors is also limited in temporal networks with given node labels. We need more work of node matching for those temporal networks without node labels. Third, consideration of all the connection series in temporal networks may cost many computational resources in dealing with very large-scale temporal networks. Finally, due to the shortage of temporal network data sets with pre-known events, our method also needs more practical validities. We plan to verify our method using more empirical temporal networks with ground truth in the future.

## 6. Data Availability

All data used in this study are publicly available on the website of http://www.sociopatterns.org/.

## 7. Conflicts of Interest

The authors declare that there is no conflict of interest regarding the publication of this paper.

## 8. Funding Statement

This work is supported by the National Science Foundation under Grant No. 1734147.

## 9. Acknowledgments

We thank Naoki Masuda for his professional suggestions and comments on this work. We thank Sociopatterns organization for selflessly disclosing their data for public use.

## 10. References

[1]    Holme, Petter, and Jari Saramäki, eds. Temporal Network Theory. Cham, Switzerland: Springer International Publishing, 2019.

[2]    Holme, Petter. "Modern temporal network theory: a colloquium." The European Physical Journal B 88.9 (2015): 234.

[3]    Li, Aming, et al. "The fundamental advantages of temporal networks." Science 358.6366 (2017): 1042-1046.

[4]    Masuda, Naoki, and Petter Holme, eds. Temporal network epidemiology. Berlin, Germany: Springer, 2017.

[5] Stehlé, Juliette, et al. "Simulation of an SEIR infectious disease model on the dynamic contact network of conference attendees." BMC medicine 9.1 (2011): 87.

[6] Perer, Adam, and Jimeng Sun. "Matrixflow: temporal network visual analytics to track symptom evolution during disease progression." AMIA annual symposium proceedings. Vol. 2012. American Medical Informatics Association, 2012.

[7] Li, Ming-Xia, et al. "Statistically validated mobile communication networks: the evolution of motifs in European and Chinese data." New Journal of Physics 16.8 (2014): 083038.

[8] Ye, Qi, et al. "Cell phone mini challenge award: Social network accuracy—exploring temporal communication in mobile call graphs." 2008 IEEE Symposium on Visual Analytics Science and Technology. IEEE, 2008.

[9] Stolfo, Salvatore J., Germán Creamer, and Shlomo Hershkop. "A temporal based forensic analysis of electronic communication." Proceedings of the 2006 international conference on Digital government research. Digital Government Society of North America, 2006.

[10] Fujiwara, Yoshi, et al. "Structure and temporal change of the credit network between banks and large firms in Japan." Economics: The Open-Access, Open-Assessment E-Journal 3 (2009): 7.

[11] Granger, Clive WJ, and Zhuanxin Ding. "Stylized facts on the temporal and distributional properties of daily data from speculative markets." UCSD Department of Economics Discussion Paper (1994): 94-19.

[12] Gao, Wei, et al. "Temporal and spatial evolution of brain network topology during the first two years of life." PloS one 6.9 (2011): e25278.

[13] Thompson, William Hedley, Per Brantefors, and Peter Fransson. "From static to temporal network theory: Applications to functional brain connectivity." Network Neuroscience 1.2 (2017): 69-99.

[14] Neal, Zachary. "The devil is in the details: Differences in air traffic networks by scale, species, and season." Social networks 38 (2014): 63-73.

[15] Sun, Xiaoqian, Sebastian Wandelt, and Florian Linke. "Temporal evolution analysis of the European air transportation system: air navigation route network and airport network." Transportmetrica B: Transport Dynamics 3.2 (2015): 153-168.

[16] Masuda, Naoki, and Petter Holme. "Detecting sequences of system states in temporal networks." Scientific Reports 9.1 (2019): 795.

[17] Dunn, Joseph C. "A fuzzy relative of the ISODATA process and its use in detecting compact well-separated clusters." Journal of Cybernetics 3.3 (1973): 32-57.

[18]     Calhoun, Vince D., et al. "The chronnectome: time-varying connectivity networks as the next frontier in fMRI data discovery." Neuron 84.2 (2014): 262-274.

[19]     Shine, James M., et al. "The dynamics of functional brain networks: integrated network states during cognitive task performance." Neuron 92.2 (2016): 544-554.

[20]     Telesford, Qawi K., et al. "Detection of functional brain network reconfiguration during task-driven cognitive states." NeuroImage 142 (2016): 198-210.

[21]     Stehlé, Juliette, et al. "High-resolution measurements of face-to-face contact patterns in a primary school." PloS one 6.8 (2011): e23176.

[22]     Isella, Lorenzo, et al. "What's in a crowd? Analysis of face-to-face behavioral networks." Journal of theoretical biology 271.1 (2011): 166-180.

[23]     Blondel, Vincent D., et al. "Fast unfolding of communities in large networks." Journal of statistical mechanics: theory and experiment 2008.10 (2008): P10008.

[24]     Goldin, Dina Q., and Paris C. Kanellakis. "On similarity queries for time-series data: constraint specification and implementation." International Conference on Principles and Practice of Constraint Programming. Springer, Berlin, Heidelberg, 1995.

[25]     Fu, Tak-chung. "A review on time series data mining." Engineering Applications of Artificial Intelligence 24.1 (2011): 164-181.

[26]     Rafiei, Davood, and Alberto Mendelzon. "Similarity-based queries for time series data." Proceedings of the 1997 ACM SIGMOD international conference on Management of data. 1997.

[27]     Newman, Mark EJ. "Detecting community structure in networks." The European Physical Journal B 38.2 (2004): 321-330.

[28]     Alvarez, Alejandro J., Carlos E. Sanz-Rodríguez, and Juan Luis Cabrera. "Weighting dissimilarities to detect communities in networks." Philosophical Transactions of the Royal Society A: Mathematical, Physical and Engineering Sciences 373.2056 (2015): 20150108.

[29]     Newman, Mark EJ. "Fast algorithm for detecting community structure in networks." Physical review E 69.6 (2004): 066133.

[30]     Koutra, Danai, et al. "DeltaCon: principled massive-graph similarity function with attribution." ACM Transactions on Knowledge Discovery from Data (TKDD) 10.3 (2016): 28.